\documentclass[aip, preprint, groupedaddress, notitlepage]{revtex4-2}
\usepackage{amsfonts} 
\usepackage{amsmath} 
\usepackage{amssymb} 
\usepackage{amsthm} 
\usepackage{mathtools} 
\mathtoolsset{showonlyrefs} 
\usepackage{slashed}  
\usepackage{tensor}

\usepackage[bookmarksnumbered,pdfpagelabels=true,plainpages=false,colorlinks=true,
            linkcolor=black,citecolor=black,urlcolor=black]{hyperref}
\usepackage{color}
\usepackage{xcolor}
\usepackage{graphicx}
\usepackage{tikz}
\usetikzlibrary{positioning}

\newtheorem{theorem}{Theorem}[section]

\newtheorem{lemma}[theorem]{Lemma}

\numberwithin{equation}{section}
\allowdisplaybreaks

\newcommand{\cD}{\mathcal{D}}

\newcommand{\cU}{\mathcal{U}}

\newcommand{\cM}{\mathcal{M}}

\newcommand{\bT}{\mathbf{T}}

\newcommand{\del}{\partial}

\newcommand{\eps}{\varepsilon}

\newcommand{\cJ}{\mathcal{J}}

\begin{document}

\title{Causal Energy-Momentum Tensors and Relativistic Fluids}
\author{V. Hoang}
\affiliation{Department of Mathematics, University of Texas at San Antonio, One UTSA Circle, San Antonio, 78249, TX, USA}

\begin{abstract}
In this paper, we consider a theory defined by an energy-momentum tensor depending on a set of general fields, including the space-time metric. We prove that if the theory is causal, bounded and transforms appropriately under diffeomorphism, it will depend only on the local values of the independent fields and their covariant derivatives up to a finite order. The implications are that the energy-momentum tensor of a causal relativistic fluid can only depend on covariant derivatives only up to a finite order.   
\end{abstract}

\maketitle

\section{Introduction}
Relativistic fluid dynamics is an effective field theory for the long-wavelength modes of a continuum system, compatible with the principles of general relativity. An essential part of any fluid description is the specification of the energy-momentum tensor $T^{ab}$, a symmetric tensor field defined on the space-time manifold. $T^{ab}$ usually depends on fields characterizing the local state of fluid, such as its temperature and four-velocity. Energy-momentum conservation is expressed by
\begin{align}
    \nabla_a T^{ab} = 0.
\end{align}
A prominent class of theories are the fluids introduced in the works of Israel, Stewart and M\"uller in Refs. \cite{MIS-1, MIS-2, MIS-3, MIS-4, MIS-5, hiscock_salmonson_1991} (see 
Refs. \cite{Baier:2007ix,Denicol:2012cn} for modern versions). These models are able to reproduce observed properties of the quark-gluon plasma (e.g. Refs. \cite{Ryu:2017qzn,Romatschke:2017ejr}). Mathematical conditions for stability, causality in the linear and nonlinear regimes have been derived in Refs. \cite{Hiscock_Lindblom_stability_1983,Olson:1989ey, BemficaDisconziNoronha_IS_bulk,DisconziBemficaHoangNoronhaRadoszNonlinearConstraints}. 

An important question is how to incorporate viscosity and other dissipative effects into the relativistic fluid equations.
A modern principled approach to this question \cite{Romatschke_2010, Loganayagam_2008,  Diles2022TwoTheorems} attempts to construct a theory by expanding the energy-momentum tensor in gradients of the fields
\begin{align}\label{eq_infinite}
    T^{ab} &= T^{ab}_{0}(\eps, u, g) + 
    T^{ab}_{1}(\del \eps, \del u, \del g) + T^{ab}_{2}(\del^2 \eps, \del^2 u, \del^2 g) + \ldots
\end{align}
The expansion \eqref{eq_infinite} is often thought to be approximate, valid for small gradients of the velocity and energy density fields, similar to a Taylor expansion. The question arises, however, if there are general principles that can be used to justify the existence of  \eqref{eq_infinite}.

In this paper, we provide such a justification. Starting from a very general point of view, we consider a theory where the energy-momentum tensor of a fluid system is a nonlinear mapping from the space of fields into the space of energy-momentum tensors. This implies in particular that non-equilibrium contributions to the energy-momentum tensor of the fluid are allowed to depend on possibly all derivatives of the fields. We will then show that principles of boundedness, locality and invariance under transformation very strongly restrict the form of this general functional: it will depend, at each point in space-time, only on a finite number of derivatives of the fields and the metric.

\section{Basic setup and notation}
Without loss of generality, we consider a fluid or any type of continuous matter system embdedded in a given $d+1$-dimensional, globally hyperbolic space-time manifold $\mathcal{M}$. On $\mathcal{M}$, we consider smooth metric tensor fields $(g_{ab})$ with Lorentzian signature $-+++\cdots$. We will also assume that $\mathcal{M}$ is time-orientable (see Ref. \cite{HawkingEllisBook}).

We consider an effective field theory described by a general energy-momentum tensor depending on a finite set of fields $\Psi = (\Psi\indices{_{(i)}^a^b^c^\ldots}(x^d))_{(i)=1, \ldots Q}$ (see e.g. \cite{HawkingEllisBook}). The fields can be scalar, vectorial or tensorial. For the case of an uncharged relativistic fluid the fields are $\Psi = (\eps, u^a, g^{ab})$ where $\eps$ is the energy density in the local rest frame of the fluid and $u^a$ the four-velocity field satisfying $u^a u_a = -1$. 

We designate a nonempty set of admissible fields for our effective field theory that we denote by $\mathcal{D}$. Elements of $\mathcal{D}$ are smooth $C^\infty$ fields $\Psi$. See Appendix A for precise definitions. We assume $\mathcal{D}$ has the \emph{homotopy property}: for any two fields $\Psi, \bar \Psi$ that coincide at some $p\in \mathcal{M}$ up to some fixed finite order $M$ in covariant derivatives, we assume that there exists a smooth homotopy $\Psi_\lambda, 0\leq \lambda \leq 1$ in $\mathcal{D}$, connecting the two pairs such that 
\begin{align}
  \Psi_0(p) = \Psi(p),~~ \Psi_1(p) = \bar \Psi(p), ~~\nabla^k \Psi_\lambda (p) = \nabla^k \Psi_0 (p)    
\end{align}
for $k \leq M$ and $\lambda\in [0,1]$. We emphasize that the metric is assumed to be one of fields in the collection $\Psi$, and when talking about covariant derivatives of the metric of order $\geq 2$, we mean derivatives of the Riemann tensor.  
See Appendix C, where the case of $\Psi = (\eps, u^a, g^{ab})$ is discussed. 

We will consider our theory to be defined by a nonlinear mapping   
\begin{align}
    \bT: \Psi \mapsto \bT[\Psi]
\end{align}
which is well-defined on $\mathcal{D}$ and maps $\Psi\in \cD$ into the fibre bundle of smooth, symmetric 2-tensor fields on $\mathcal{M}$. This means that in any local coordinate system, the components of the tensor field $T^{ab}[\Psi]$ depend in a completely general, possibly nonlinear way on the fields. 

\begin{figure}
    \centering
    \includegraphics[scale=0.5]{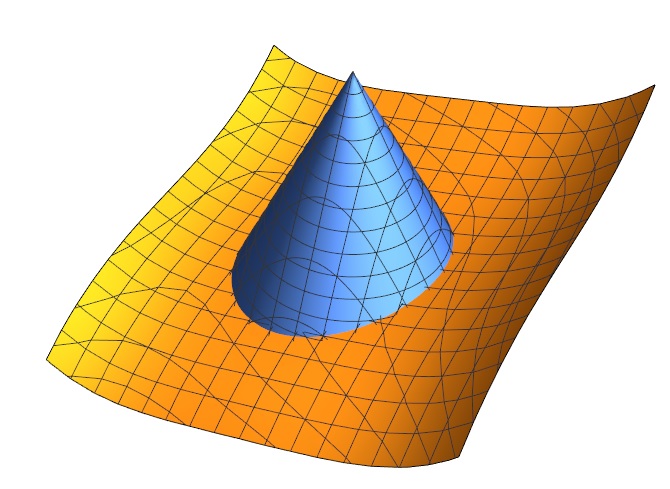}
    \caption{Backwards light-cone terminating on a space-like hypersurface to its past.}
    \label{fig1}
\end{figure}

Next, we assume that the variation $\del \bT$ of $\bT$ exists and is a well-defined operator mapping $C^\infty$ fields into $C^\infty$ fields:
\begin{align}
    \del\bT[\Psi](\Phi) = \lim_{h\to 0} \frac{1}{h}(\bT[\Psi + h \Phi] - \bT[\Psi])
\end{align}
and that $\del \bT[\Psi]$ depends continuously on $\Psi$. 
We use $\cJ^{-}_{\Sigma}(p)$ to denote the backwards light-cone with apex $p\in \mathcal{M}$ and terminating on $\Sigma$, a space-like hypersurface in the past of $p$. See Figure \ref{fig1} for an illustration and see also Ref. \cite{HawkingEllisBook}.

The postulates for $\bT$ are:

\begin{enumerate}
\item[(C)] \emph{\underline{C}ausality}. If $p$ is any point in space-time and let $\Psi, \bar \Psi\in \mathcal{D}$. If there exists a $\Sigma$ such that 
\begin{align}
        \Psi = \bar \Psi\quad \text{in $\cJ^{-}_\Sigma(p)$}
\end{align}
 then 
\begin{align}
    \bT[\Psi]|_p = \bT[\bar \Psi]|_p ~~~\text{at the point $p$}.
\end{align}

\item[(B)] \emph{\underline{B}oundedness of variation.} For any open set $\cU\subset \cM$ with compact closure, $p\in \cU$ and any fixed $C^\infty_c$ field $\Phi$ with support inside $\cU$
\begin{align}
    \sup_{\Psi\in \mathcal{D}, p\in \cU} |\del \bT[\Psi](\Phi)|_p| < \infty.
\end{align}
Here, boundedness is defined in terms of the $C^\infty$ topology, see Appendix A. The quantity $|\del \bT[\Psi](\Phi)|_p|$ is defined as
\begin{align}
    |\del \bT[\Psi](\Phi)|_p| = \max\{|\lambda|\}
\end{align}
where $\lambda$ are the eigenvalues of $T^a_b[\Psi](\Phi)|_p$.
\item[(T)] \emph{Universality under \underline{T}ransformations.} For any diffeomorphism $\phi: \mathcal{M} \to \mathcal{M}$, we have
\begin{align}
    \bT[\phi^{*} \Psi] = \phi^{*} \bT[\Psi]
\end{align}
Here, $\phi^*$ is the push-forward of tensor fields. 
This requirement simply states that our theory is diffeomorphism invariant. 
\end{enumerate}

The push-forward is defined as follows: In local coordinates, let the diffeomorphism $\phi: \cU_1 \to \cU_2$ be described in components by
$\phi^a(x^c), a,c = 0\ldots d$. Then
\begin{align}
    &\phi^* \Psi^K|_{\phi(p)} = 
    \del_{k_1'} \phi^{k_1}(p) \ldots \del_{k_n'} \phi^{k_n}(p)  \Psi^{K'}|_{p} 
\end{align}
where $K=(k_1, \ldots), K' = (k_1', \ldots)$.
Moreover, a direct consequence of (T) is that
\begin{align}\label{eq_transform_linearization}
    \del \bT[\phi^{*} \Psi](\phi^{*} \Phi) = \phi^{*} [\del \bT[\Psi](\Phi)].
\end{align}

An important consequence of (C) is that the variation of $T^{ab}$ is causal: 
\begin{align}\label{apex_zero}
    \text{$\Phi(x) = 0$ in some $\cJ^{-}_\Sigma(p)$} \Rightarrow \del \bT[\Psi](\Phi)|_{p} =  0.
\end{align}

We shall prove the following Theorem. 
\begin{theorem}\label{main}
Assuming postulates (C), (B) and (T): there exists an integer $M \geq 0$ such that the energy-momentum tensor at each space-time point $p$ only depends on the physical fields $\Psi$ and their covariant derivatives up to order $M$ at that the same space-time point $p$:
\begin{align}
    \bT[\Psi]|_p = \bT(\Psi, \nabla \Psi, \ldots, \nabla^M \Psi)|_{p}
\end{align}
\end{theorem}.

\section{Proof of Main result}
Let two fields $\Psi, \bar \Psi\in \mathcal{D}$ be given and let $p$ be a fixed point in space-time. We will show that there exist a positive integer $M$, independent of $\Psi, \bar \Psi$ and $p$ such that if all the covariant derivatives of $\Psi$ and $\bar \Psi$ coincide up to order $M$ at $p$, then $\bT[\Psi] = \bT[\bar \Psi]$ at $p$. Choose a local system of coordinates valid in a neighborhood $\cU$ of $p$. We use $\Psi^K, \bar\Psi^K, T^{ab}$ for the components of the tensors involved. By the homotopy property, there exists a smooth homotopy $s\in [0,1] \mapsto \Psi_s \in \cD$ such that $\Psi_0 = \Psi, \Psi_1 = \bar \Psi$. We then have
\begin{align}\label{eq_integral}
    &T^{ab}[\bar \Psi] - T^{ab}[\Psi] = \int_0^1 \del T^{ab}[\Psi_s](\Psi'_s) ds
\end{align}
where $\Psi'_s = \frac{d}{ds}\Psi_s$. 

In Appendix B, we show the following Lemma:
\begin{lemma} Let (C), (B), (T) hold. There exists an integer $M\geq 0$ such that the variation $\del \bT[\Psi](\Phi)$ is a differential operator of order $M$ with smooth coefficients acting on the fields $\Phi=(\Phi_{(1)}, \Phi_{(2),}\ldots)$. More precisely, in any coordinate neighborhood of $p\in \mathcal{M}$, $\del \bT$ can be written as 
\begin{align}\label{eq_local_rep}
    \del T^{ab}[\Psi](\Phi) = \sum_{(i)}\sum_{|J|\leq M} c^{ab (i) J}_{K}[\Psi] ~\del_J \Phi^K_{(i)}.
\end{align}
\end{lemma}
Here, we use uppercase Latin letters to denote multiple contravariant or covariant indices; thus a tensor of type $(r, s)$ will be written as $\psi^I_J$ and we denote by $|I| = r$ the number
of indices that the multiple index $I$ represents. 

Now suppose that for $\Psi, \bar \Psi$ we have that all covariant derivatives of $\Psi, \bar \Psi$ of order less than $M$ coincide at $p$. We choose a homotopy $s\mapsto \Psi_s$ such that $\nabla^{k}\Psi_s = \nabla^{k}\Psi = \nabla^{k}\bar \Psi$ at $p$ for all $s\in [0,1], k\leq M$. Hence all derivatives of order less than $M$ of $\Psi'_s$ vanish at the point $p$. 

This implies that $\del T^{ab}[\Psi(s)](\Psi'_s)|_p = 0$ and by \eqref{eq_integral}, we have
$T^{ab}[\Psi]|_p = T^{ab}[\bar \Psi]|_p$. This means that $T^{ab}[\Psi]|_p$ only depends on derivatives of at most order $M$ at $p$. 

\section{Conclusion}
In this paper, we have studied consequences of causality and shown that under very general assumption, a causal energy-momentum tensor can only depend on covariant derivatives of the underlying independent fields up to a certain finite order. 

\section{Appendix}

\subsection{Appendix A: Topology of $C^\infty$}
We recall some facts about the topology of $C^\infty$ (see Ref. \cite{Narasimhan_1985}), which make it possible to define convergence of sequences. Let $\Omega$ be a nonempty, open subset of the space-time manifold. Let $K_\nu, \nu=0, 1, 2\ldots$ be a sequence of open subsets of $\Omega$ such that the closure $\bar K_\nu$ is compact and such that $\bar K_\nu \subset K_{\nu+1}$, the union of all the $K_\nu$ being $\Omega$. The topology of $C^\infty(\Omega)$ is induced by the following metric
\begin{align}
    d(f, g) = \sum_{\nu=0}^\infty 2^{-\nu}\frac{\|f- g\|_{\nu, K_\nu}}{1+\| f - g \|_{\nu, K_\nu}}
\end{align}
where $\|f\|_{\nu, K_\nu} = \sup_{x\in K_\nu, |\alpha|\leq \nu} |D^{\alpha} f(x)|$, i.e. the supremum norm of derivatives of $f\in C^\infty(\Omega)$ of order $\leq \nu$. For more information, see Ref. \cite{Narasimhan_1985}, Chapter 1.

The following result, Peetre's theorem \cite{Narasimhan_1985}, plays an important role:
\begin{theorem}
Let $\xi, \eta$ be two vector bundles over the space-time manifold. $\Lambda : C^\infty(\Omega, \xi) \to C^\infty(\Omega, \eta)$ be a linear mapping from the space of sections of the two vector bundles with the following property:
\begin{align}\label{property_support}
    \operatorname{supp}{\Lambda f} \subseteq \operatorname{supp}f 
\end{align}
for any $f\in C^\infty(\Omega, \xi)$. Then there exists a positive integer $M$ such that the operator $\Lambda$ can be written as
\begin{align}
    \Lambda = \sum_{|\alpha|\leq M} a_\alpha(x) D^\alpha \label{eq::Peetre_representation}
\end{align}
with smooth (possibly matrix-valued) coefficients $a_\alpha$.
\end{theorem}
The action \eqref{eq::Peetre_representation} is to be understood locally: in a neigborhood of any $p\in \Omega$, there exist local trivializations of the fibre bundles, i.e. sections $f$ are represented in local frames by component functions (see Ref. \cite{Narasimhan_1985}). Note that Peetre's theorem only applies to linear mappings.  

\eqref{property_support} implies 
\begin{align}\label{property_support1}
    (\Lambda (f + g))(x) = (\Lambda f)(x)
\end{align}
if $x$ is not in the support of $g$. 

Next we show that $\del \bT[\Psi]$ possesses the property \eqref{property_support}. This is seen as follows: Let $x$ be a point in the complement of the support of $\Phi$, i.e.~$\Phi = 0$ in an open neigborhood $\cU_x$ of $x$. Each point $y\in \cU_x$ is the apex of a small $J_{\Sigma}^-(y)$ contained in $\cU_x$. Therefore by \eqref{apex_zero}, $\del \bT[\Psi](\Phi)|_{y} = 0$ for any $y\in \cU_x$. 
It follows that $x$ is not contained in the support of $\del \bT[\Psi](\Phi)$. 

\subsection{Appendix B: Proof of Lemma}
We will study the linearization $L := \del \bT[\Psi]$ and show that it is a differential operator of finite order $M$, the number $M$ not depending on $\Psi$. We fix a point $p\in \cM$, an open neighborhood of $\cU$ of $p$ and pass to local coordinates with $p$ being the origin of the coordinate system. Let $B_r(p)$ be the coordinate ball with (small) euclidean radius $r > 0$ with center $p$. The definition of $B_r(p)$ depends on the specific coordinate system. 

We claim: \emph{There exists a constant $C > 0$, an integer $M\geq 0$ and a $r > 0$ such that for all $\Psi \in \mathcal{D}$}
\begin{align}
    \|L[\Psi](\Phi)\|_{0, B_r(p)} \leq C \|\Phi\|_{M, B_r(p)}\label{ineq_main}
\end{align}
\emph{holds for any $\Phi$ supported in a ball of radius $r>0$ around $p$ and such that $\Phi(p) = 0$.}  Assume by contradiction that the statement is not true. Then for any $C>0$, $M\geq 0$, any $r>0$ there exists a $\Psi$ and a $\Phi$ supported in $B_p(r)$ and vanishing at $p$, such that
\begin{align}\label{eq_contradict}
    \|L[\Psi](\Phi)\|_{0, B_p(r)} > C \|\Phi\|_{M, B_p(r)}.    
\end{align}
We claim that it is possible to inductively construct a sequence $r_k>0,\Psi_k, \Phi_k$ such that 
\begin{align}
    \|L[\Psi_k](\Phi_k)\|_{0, B_{r_k}(p)} > 2^{2k} \|\Phi_k\|_{k, B_{r_k}(p)} \label{eq:proof_lemma_eq1}  
\end{align}
holds and such that the support of $\Phi_{k+1}$ does not intersect the support of $\Phi_{k}$. Namely, after $r_k, \Psi_k, \Phi_k$ are known, we pick $r_{k+1}>0$ small enough such that $B_{r_{k+1}}(p)$ does not intersect the support of $\Phi_{k}$. This is possible since $\Phi_{k}(p) = 0$. We then apply \eqref{eq_contradict} to find $\Psi_{k+1}, \Phi_{k+1}$. 

Define the $C^\infty$ field $\Phi$ via the following series
\begin{align}\label{eq_series}
    \Phi := \sum_{k=1}^\infty 2^{-k} \|\Phi_k\|_{k, B_{r_k}(p)}^{-1} \Phi_k
\end{align}
Note that \eqref{eq:proof_lemma_eq1} implies that $\|\Phi_k\|_{k, B_{r_k}(p)} > 0$. 
By the \eqref{property_support1} we have 
\begin{align}
    L[\Psi_k](\Phi)= 2^{-k} \|\Phi_k\|_{k, B_{r_k}(p)}^{-1} L[\Psi_k](\Phi_k)~~~\text{on $B_{r_k}(p)\setminus B_{r_{k+1}}(p)$}
\end{align}
and therefore
\begin{align}
    \|L[\Psi_{k}](\Phi)\|_{0, B_{r_k}(p)} > 2^k \to \infty 
\end{align}
as $k\to \infty$. This is a contradiction to postulate (B), and hence \eqref{ineq_main} holds. We now intend to apply Peetre's theorem to the mapping $L$.  Hence there exists an $N = N(\psi)$ such that 
\begin{align}\label{eq_rep_peetre}
    L^{ab}[\Psi](\Phi) = \sum_{(i)}\sum_{|J| \leq N(\Psi)} c^{ab J}_{K} [\Psi] \del_J \Phi^K_{(i)}.
\end{align}
Note that the order $N(\Psi)$ of the differential operator $L[\Psi]$ may initially depend on $\Psi$. However, the inequality \eqref{ineq_main} implies that on $B_r(x)\setminus\{x\}$, $L^{ab}[\Psi]$ has an order less than $M$ for any $\Psi\in \mathcal{D}$. This is shown in 3.3.6 of Reference \cite{Narasimhan_1985}. Since the coefficients of the differential operator are smooth, we conclude $N(\psi)\leq M$ on the whole ball $B_r(p)$.  

To summarize, we have shown for any fixed $p$, there exists a representation of the form \eqref{eq_local_rep} in a neighborhood of $p$. The order of the differential operator is $M=M(p)$ and at this stage may depend on $p$. Our goal is to remove this dependency on $p$ and have a uniform order $M$ for the whole space-time manifold. To finish the proof, we therefore need to use property (T). Take any other $\tilde p\in \mathcal{M}$. There exists a diffeomorphism $\phi: \cM \rightarrow \cM$ such that $\phi(\cU) = B_r(p)$, where $\mathcal{U}$ is some neigborhood of $\tilde p$. By (T), 
\begin{align}\label{order_M_1}
     \del \bT [\Psi](\Phi) =  \phi_{*} \del \bT[\phi^{*} \Psi](\phi^{*}\Phi) 
\end{align}
where $\phi_{*}$ is the pullback to $\cU$. $\bT[\phi^{*} \Psi](\phi^{*}\Phi)$ on the right-hand side of \eqref{order_M_1} is a differential operator of order $M$, and since application of $\phi_{*}$ does not change the order, $\del \bT [\Psi](\Phi)$ acts as a diferential operator of order $M$ in $\cU$ as well.

\subsection{Homotopy of fields}
Here, we describe the construction of $\cD$ for the simplest model for an uncharged relativistic fluid depending on the energy density, space-time metric and four-velocities, i.e.~the fields $\Psi = (\eps, g^{ab}, u^a)$. In order to apply Theorem \ref{main}, we have to make sure that two $\Psi, \bar \Psi$ that coincide up to derivative order $M$ at a point $p$ can be smoothly connected by a homotopy $\lambda \mapsto \Psi_\lambda$ with $\Psi_0 = \Psi$ and $\Psi_1 = \bar \Psi$. Moreover, $\Psi_\lambda \in \cD$ should hold for $\lambda \in [0,1]$, i.e. the homotopy should not leave the predefined set of \emph{physical fields} $\cD$. While we have a certain leeway in choosing $\cD$, a reasonable choice is as follows: 
\begin{enumerate}
    \item[(a)] For $\eps$, we take scalar fields with $\eps(x) > 0$. The homotopy between fields $\eps, \bar \eps$ is defined by 
    \begin{align}
        \eps_\lambda = \lambda \eps + (1-\lambda) \bar \eps.
    \end{align}
    Note that $\eps_\lambda > 0$. 
    \item[(b)] We represent each metric of physical interest by a smooth covector frame $\{\theta_{a}^{(i)}(x^c)\}_{i=0, 1, 2, 3}$. The metric tensor is then defined by
    \begin{align}\label{def_g}
        g_{ab} = - \theta_{a}^{(0)}\theta_{b}^{(0)} + \sum_{i=1}^3 \theta_{a}^{(i)}\theta_{b}^{(i)}.
    \end{align}
    We claim that each pair $(\theta, \bar \theta)$ co-vector frames can be connected by a smooth homotopy $\theta_\lambda$ while keeping the property that the covectors in the frame remain linearly independent. In a neighborhood of $p$, this can simply be done by defining $\theta_\lambda = \lambda \theta + (1-\lambda)\bar \theta$ for each individual covector $\theta$. The homotopy may have to be appropriately modified on the remaining manifold, to ensure that $\{\theta_\lambda\}$ form a frame at each point of $\cM$. The corresponding metrics  defined by \eqref{def_g} are connected by the homotopy  
    \begin{align}
         (g_\lambda)_{ab} = - (\theta_\lambda)_{a}^{(0)}(\theta_\lambda)_{b}^{(0)} + \sum_{i=1}^3 (\theta_\lambda)_{a}^{(i)}(\theta_\lambda)_{b}^{(i)}.
    \end{align}
    \item[(c)] The velocity fields are all time-like, future pointing, normalized fields $u^a$ with $u^a u_a = -1$. The homotopy is defined by
    \begin{align}
        u_\lambda^c =  [- \tilde{u}^a_{\lambda} \tilde u_{\lambda,a}]^{-1/2} \tilde u^c_{\lambda}
    \end{align}
    where $\tilde u^c_\lambda = \lambda u^a + (1-\lambda) \bar u^c$.
\end{enumerate}

\bibliography{References.bib}

\end{document}